\documentclass[%
 reprint,
 amsmath,amssymb,
 aps,
]{revtex4-2}

\usepackage{graphicx}% Include figure files
\usepackage{dcolumn}% Align table columns on decimal point
\usepackage{bm}% bold math
\usepackage{xcolor}
\usepackage{etoolbox}
\usepackage{comment}
\begin{document}

\preprint{APS/123-QED}

\title{Non-local optical response of a multi-phased quantum material}% Force line breaks with \\
% \thanks{A footnote to the article title}%

\author{D. Zhang}
 \altaffiliation{Department of Physics and Astronomy, Rice University, 6100 Main Street, Houston, TX 77005}
\author{G. V. Naik}%
 \email{guru@rice.edu}
\affiliation{%
 Electrical and Computer Engineering Department, Rice University, 6100 Main Street, Houston, TX 77005}%

\date{\today}% It is always \today, today,
             %  but any date may be explicitly specified

\begin{abstract}
Light-matter interaction in quantum materials presents a new paradigm as light can tip the balance between many competing quantum many-body phases to result in new phenomena. Describing the optical response of such materials requires complex models. Here, we develop a non-local model to describe the optical response of a quantum material, 1T-TaS$_2$. 1T-TaS$_2$ is a nearly commensurate charge-density-wave material at room temperature. The competing stacking configurations of the charge domains in this layered material result in significant optical inhomogeneity that necessitates a non-local dielectric function. We experimentally measure the non-local optical response of 1T-TaS$_2$ films under various illumination intensities and validate our model. The non-local parameter extracted from our measurements sheds light on the competition between the two stacking configurations of 1T-TaS$_2$. Our technique of measuring non-local optical response serves as a quick, simple, and non-invasive method to probe the energy landscape of strong correlations in many such quantum materials.
\end{abstract}

\maketitle

\section{\label{sec:level1}Introduction}
Quantum materials are unique in their long-range interactions and competing phases tunable by external stimuli. Light makes an excellent stimulus and probe to study the energy landscape of quantum materials~\cite{scalari2012ultrastrong,wurstbauer2017light,zengin2015realizing,dufferwiel2015exciton}. One of the important questions that arise in this context is whether the optical response of quantum materials can be described by a dielectric function. Here, we address this question and show that the dielectric function has to be non-local to accurately describe the light-matter interaction. Due to the incommensuracy of the quantum order or competing phases, the volume of quantum materials is partitioned into multiple domains~\cite{loret2019intimate,borisenko2008pseudogap}. This electronic inhomogeneity in the material results in a non-local optical response~\cite{elser2007nonlocal,eggebrecht2017light}. Here, we study such non-local optical response of 1T-TaS$_2$, a layered quantum material supporting charge density waves (CDW) at room temperature~\cite{sipos2008mott,manzke1989phase,gruner1985charge,gruner1988dynamics}. 

\begin{figure*}
\centering
\includegraphics{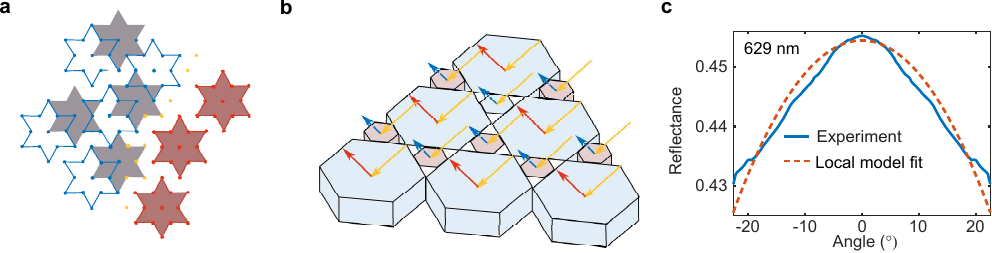}
\caption{\label{fig:panel1} (a). Stacking of CDW domains in 1T-TaS$_2$: The green dots represent Ta atoms in the ab-plane. David stars are the basic CDW units. Stars in the bottom and top layers are represented by gray and red shades. Two types of stacking orders are illustrated: center-to-center or A-stacking denoted by overlapping red David stars and center-to-corner or L-stacking denoted by partially overlapping red and blue stars. (b). An illustration of the arrangement of domains with different stacking orders (red for A and blue for L) at the surface of 1T-TaS$_2$ (c). Measured and calculated reflectance curves at 629 nm wavelength for a thin film of 1T-TaS$_2$ for various angles of incidence. The calculations used the best-fit local dielectric function for 1T-TaS$_2$.}
\end{figure*}

CDW in the 1T-TaS$_2$ manifests as picometer-dislocations in the lattice to form David star-like structures~\cite{thomson1994scanning} as depicted in Fig.~\ref{fig:panel1}a. Such stars hosting one electronic charge fill every layer of this material. At room temperature, the stars do not fill the entire layer but leave metallic gaps at the domain walls~\cite{yamada1977origin,wen2019photoinduced,park2021zoology}.

%Emergent sub-orders transpire along the domain walls and in the bulk of domains~\cite{lantz2017domain}. 

The David stars not only interact with other stars in the same layer but also weakly interact with stars in neighboring layers (along the c-axis). This c-axis interaction results in various configurations of stacking of the stars. Center-to-center (A) and center-to-corner (L) configurations~\cite{ritschel2018stacking,stahl2020collapse,butler2020mottness} are the two major stacking orders, as shown in Fig.~1(a). These two stacking configurations co-exist and compete~\cite{li2021light}. As a result, a light-tunable or non-linear optical response was observed at low-intensity illumination~\cite{li2020large}. Also, such tunable optical response arising from the same competing stacking orders was observed in 1T-TaS$_2$ films under an in-plane electrical bais~\cite{li2019plane,li2021reorganization}. All these studies showed that the optical responses of A-stacking and L-stacking orders are different. Fig.~1(b) illustrates it by showing the domains with different stacking orders (red for A and blue for L) resulting in different polarization responses. Past studies homogenized the net optical response of the film based on the fill fractions of domains with A and L stacking and assigned a local dielectric function \cite{li2019plane,li2020large,li2021light,li2021reorganization}. 
%Here, we question the accuracy of such a local model for the dielectric function of a spatially inhomogeneous material. 
%The A domains exist in the bulk region, while the L domains exist near the domain walls~\cite{li2021reorganization}.
However, when the domain sizes are not too small compared to the wavelength of light, homogenization of optical response cannot accurately describe the light-matter interaction. Therefore, a non-local dielectric function is needed~\cite{elser2007nonlocal}. Fig.~1(c) plots the measured angle-dependent reflectance of a 1T-TaS$_2$ film in p-polarization. The best local dielectric function fit to the experimental data is still quite inaccurate, especially at higher angles. This error, about 0.7\% -- 1\%, exceeds the error of our well-calibrated measurement system by twice or more. Thus, an angle-dependent c-axis permittivity or non-local dielectric function is needed to better describe the optical response of 1T-TaS$_2$.  

In the following sections, we present the derivation of a non-local dielectric model based on the effective medium theory, apply this model to 1T-TaS$_2$, and compare the calculated and experimental reflectance data. Further, we show that the non-local parameter in our model provides valuable information on domain size and its dependence on illumination intensity.  

\section{Non-local optical response model}

\begin{figure*}
\centering
\includegraphics{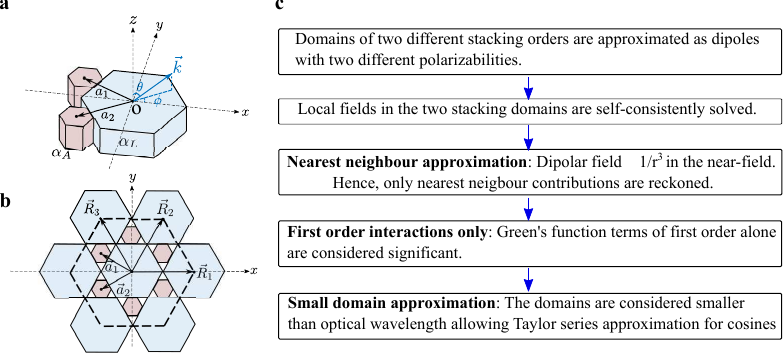}
\caption{\label{fig:panel2} (a) Schematic of a unit cell showing domains of A-stacking (red) and L-stacking (blue). The reference coordinate system and the direction of light are shown here. (b) Close packing arrangement of domains with A and L stacking orders (red and blue, respectively). A unit cell for effective susceptibility calculations is highlighted in dashed lines. (c) A flow chart highlighting key steps in the derivation of non-local dielectric function for 1T-TaS$_2$.}
\end{figure*}

The effective medium theory (EMT) allows homogenizing the optical response of an inhomogeneous medium. The EMT works well only if the inhomogeneity length scale, $a<<\lambda/2\pi$, where $\lambda$ is the free space wavelength of light. The stacking domain size in 1T-TaS$_2$ at room temperatures is comparable to $\lambda/2\pi$ in the visible range. Hence, a non-local optical model is necessary to accurately describe the optical response of 1T-TaS$_2$ at room temperature. 

Our non-local model assumes a periodic arrangement of the A and L stacking domains in the top layer of domains. A nonperiodic arrangement would not significantly alter the results as long as the average volume fractions of the domains remain the same. Since 1T-TaS$_2$ is highly absorbing in the visible, we can safely disregard the interaction of light with the domains underneath the top layer of domains \cite{li2020large}. At room temperature, the stacking domains are about 30 nm thick and the skin depth of visible light is also about the same~\cite{lantz2017domain,li2019plane}. The two types of stacking domains, A and L, in the top layer, have different volume fill fractions, $f_{A}$ and $f_{L}$. We assume that the polarization of a domain may be represented by a point dipole located at the center of the domain with a moment $p$ that equals the product of polarizability and the local electric field. The molecular polarizabilities of the A and L stacking domains are represented by $\alpha_A$ and $\alpha_L$, respectively. The frequency dependency of the molecular polarizability functions is assumed Lorentzian $\mathcal{L}(\omega)$. We shy away from using other oscillator models because the broad resonance in the near-visible to visible range in metallic transition metal dichalcogenides like TaS$_2$ are better described by a sum of Lorentz oscillators~\cite{munkhbat2022optical}. Thus, we assume that $\alpha_{L(A)}=\epsilon_0 \alpha_0 V_{L(A)} \mathcal{L}_{L(A)}(\omega)$ where $epsilon_0$ is free space permittivity, $\alpha_0$ is a dimensionless constant, and $V_{L(A)}$ is the volume of the respective domain in a unit cell. In our measurement range, one Lorentz oscillator for each stacking type is adequate for describing the c-axis optical response.

Our approach to self-consistently solve for non-local susceptibility is shown in Fig.~\ref{fig:panel2}c. The key difference between our model and other effective medium theories is that we do not homogenize the local electric fields in the two types of domains. Instead, we represent them as $E_A$ and $E_L$ and solve the equations self-consistently to determine the fields. 

Note that the c-axis or z-direction optical properties are the focus of this work. Hence, the incident field $\vec{E}_0$ is considered polarized along the c-axis (z-axis in our model) and propagating in the $x-z$ plane: $\vec{E}_0=\hat{z}E_0e^{ik_xx+ik_zz}$. The uniaxial anisotropy of the material in the chosen coordinate system guarantees that fields in $z$-direction do not mix with those in $x$ and $y$ directions. Hence, we consider only the $z$-component of the field and $zz$-component of all tensors here.

The electric field ansatz for our periodic system for the zeroth order is given by Bloch's theorem as $E=Fe^{i\vec{k}\cdot\vec{R}}$. Here, $\vec{k}=k_x\hat{x}+k_z\hat{z}$, $\vec{R}$ is the point vector in the $x-y$ plane, $F$ is a periodic function in the $x-y$ plane with the same period as the lattice. Higher orders of diffraction are not considered here because, in the 1T-TaS$_2$ case, the lattice period is smaller than the free space wavelength of incident light.

Note that our periodic system is a lattice with three bases, the domain of stacking type L and two adjacent domains of stacking type A. We chose a coordinate system such that the centers of domains with stacking type L form the lattice points (see Fig.~\ref{fig:panel2}a). These points are denoted by lattice vectors $\vec{R}_n$. The centers of two stacking type-A domains in the unit cell are displaced from the lattice center by $\vec{a}_{1}$ and $\vec{a}_{2}$. We denote the $F(\vec{R}_n)=F_L$ and $F(\vec{R}_n+\vec{a}_{1(2)})=F_A$. Here, we assume that the electric field solutions of the two distinguishable stacking A domains are the same. Using this notation, $E(\vec{R}_n)=F_Le^{i\vec{k}.\vec{R}_n}$ and $E(\vec{R}_n+\vec{a}_{1(2)})=F_Ae^{i\vec{k}.(\vec{R}_n+\vec{a}_{1(2)})}$. 

Since the local electric field is a sum of the incident electric field and the fields from the polarized dipoles, the local electric field at the domain centers may be self-consistently solved from Eq.~\ref{eq3}. 
% The derivation of Eq.~\ref{eq3} is given by Appendix Eqs.~\ref{appendix:eq1} and \ref{appendix:eq2}. 

% \begin{align}
% \begin{split}
% E_1 = &E_0+\omega^2\mu_0\sum_{n_1\neq 0}\stackrel{\leftrightarrow}{G}_{zz}(\vec{R}_{n1})\alpha_1 F_1 e^{i\vec{k}\cdot\vec{R}_{n1}}\\
% &+\omega^2\mu_0\sum_{n_2}\stackrel{\leftrightarrow}{G}_{zz}(\vec{R}_{n2})\alpha_2 F_2 e^{i\vec{k}\cdot\vec{R}_{n2}}
% \end{split}
% \label{eq2}
% \end{align}

% \begin{align}
% \begin{split}
% E_2 = &E_0 e^{i\vec{k}\cdot\vec{r}_2}\\
% &+\omega^2\mu_0\sum_{n_1}\stackrel{\leftrightarrow}{G}_{zz}(\vec{R}_{n1}-\vec{r}_2)\alpha_1 F_1 e^{i\vec{k}\cdot(\vec{R}_{n1}-\vec{r}_2)}\\
% &+\omega^2\mu_0\sum_{n_2\neq0}\stackrel{\leftrightarrow}{G}_{zz}(\vec{R}_{n2}-\vec{r}_2)\alpha_2 F_2 e^{i\vec{k}\cdot(\vec{R}_{n2}-\vec{r}_2)}
% \end{split}
% \label{eq3}
% \end{align}

% As both F$_L$ and F$_A$ consist of dipolar fields from the same type of stacking order and the opposite type of stacking order, this leads to a 2 by 2 matrix equation: 
\begin{align}
    \begin{bmatrix}
        F_L\\
        F_A
    \end{bmatrix} =&  \begin{bmatrix}
        E_0\\
        E_0
    \end{bmatrix}
    +
    \omega^2\mu_0    
\begin{bmatrix}
        g_{LL} & g_{LA}\\
        g_{AL} & g_{AA}
    \end{bmatrix}
    \begin{bmatrix}
        \alpha_L & 0\\
        0 & \alpha_A
    \end{bmatrix}
    \begin{bmatrix}
        F_L\\
        F_A
    \end{bmatrix}
    \label{eq3}
\end{align}

\noindent where $\omega$ is the angular frequency of light, and $\mu_0$ is the permeability of free space. g factors are defined by: 
\begin{align}
g_{LL}=\sum_{n\neq0}&\stackrel{\leftrightarrow}{G}_{zz}(\vec{R}_{n}) e^{i\vec{k}\cdot\vec{R}_{n}}\\
g_{LA}=\sum_{n}&\stackrel{\leftrightarrow}{G}_{zz}(\vec{R}_{n}+\vec{a}_1) e^{i\vec{k}\cdot(\vec{R}_{n}+\vec{a}_1)}+\nonumber \\
&\stackrel{\leftrightarrow}{G}_{zz}(\vec{R}_{n}+\vec{a}_2) e^{i\vec{k}\cdot(\vec{R}_{n}+\vec{a}_2)} \\
g_{AL}=\sum_{n}&\frac{1}{2}\Big(\stackrel{\leftrightarrow}{G}_{zz}(\vec{R}_{n}-\vec{a}_1)e^{i\vec{k}\cdot(\vec{R}_{n}-\vec{a}_1)}+\nonumber \\
&\stackrel{\leftrightarrow}{G}_{zz}(\vec{R}_{n}-\vec{a}_2)e^{i\vec{k}\cdot(\vec{R}_{n}-\vec{a}_2)}\Big) \\
\begin{split}
g_{AA} = \sum_{n}&\frac{1}{2}\Big(\stackrel{\leftrightarrow}{G}_{zz}(\vec{R}_{n}+\vec{a}_2-\vec{a}_1) e^{i\vec{k}\cdot(\vec{R}_{n}+\vec{a}_2-\vec{a}_1)}+ \\
&\stackrel{\leftrightarrow}{G}_{zz}(\vec{R}_{n}+\vec{a}_1-\vec{a}_2) e^{i\vec{k}\cdot(\vec{R}_{n}+\vec{a}_1-\vec{a}_2)}\Big)+\\
\sum_{n}&\stackrel{\leftrightarrow}{G}_{zz}(\vec{R}_{n})e^{i\vec{k}\cdot\vec{R}_{n}}
\end{split}
\label{eq4}    
\end{align}
here, $\stackrel{\leftrightarrow}{G}_{zz}(\vec{r}-\vec{r'})$ is the $zz$-component of the dyadic Green's function~\cite{novotny2012principles}.

Solving Eq.~\ref{eq3} provides the local electric field strength at the centers of the A and L domains. Using the electric field strength, polarization may be computed. Polarization would lead to an effective susceptibility. However, solving Eq.~\ref{eq3} is not easy due to the summations in Eqs.~2-5 running over all domains. A simplification could be made by noticing that the magnitude of $\stackrel{\leftrightarrow}{G}_{zz}(r)$ drops as $1/r^3$ in the vicinity of the dipole, and $g$ could be approximated to sum over only the nearest neighbors. As the dipoles are modeled to exist on the same $x-y$ plane, and the first-ordered neighbors are located at the same distance, Green's function terms $G_{zz}$ are the same in each summation, respectively, in Eqs.~2-5. With hexagonal symmetry from the underlying 1T structure, the complex exponentials add up to cosines to uphold time-reversal symmetry.    

The effective susceptibility $\chi_{\text{eff}}$ of the medium is obtained by averaging the dipole moment of each stacking domain over the volume of the unit cell $V_{\text{unit}}$. Noticing the symmetry of the lattice, the unit cell is chosen as illustrated in Fig.~\ref{fig:panel2}b. $\chi_{\text{eff}}$ is thus constructed from solutions of the amplitude of the electric fields, $F_A$ and $F_L$. To simplify the closed-form expression of $\chi_{\text{eff}}$, we define two functions, domain susceptibility functions, $\chi_{L(A)}$ as $\chi_{L(A)} = \alpha_{L(A)}F_{L(A)}/V_{L(A)}\epsilon_0$. %where $\epsilon_0$ is the vacuum permittivity and $V_{L(A)}$ is the volume of an individual domain. 
Then, $\chi_{\text{eff}}$ is given by Eq.~\ref{eq5}.

\begin{align}
\begin{split}
    \chi_{\text{eff}} = &\frac{V_L}{V_{\text{unit}}}\chi_L\Big(1+\frac{2\,\text{cos}(\vec{k}\cdot\vec{R}_1)}{3}+\frac{2\,\text{cos}(\vec{k}\cdot\vec{R}_2)}{3}+ \\ 
    &\frac{2\,\text{cos}(\vec{k}\cdot\vec{R}_3)}{3}\Big) + \frac{V_A}{V_{\text{unit}}}\chi_A\Big(2\,\text{cos}(\vec{k}\cdot\vec{a}_1)+\\
    &2\,\text{cos}(\vec{k}\cdot\vec{a}_2)+ 2\,\text{cos}(\vec{k}\cdot(\vec{a}_1-\vec{R}_2))\Big)
\end{split}
    \label{eq5}
\end{align}

% The cosines result from summations over complex exponential phases with the symmetry shown in Fig.~\ref{fig:panel2}b. $\chi_{L(A)}$ are obtained from solving Eq.~\ref{eq3}. 
Using the definition of the polarizability $\alpha_{L(A)}=\epsilon_0\alpha_0 V_{L(A)} \mathcal{L}_{L(A)}$, domain susceptibility functions may be solved to the expressions shown in Eq.~\ref{eq7}.

\begin{align}
\begin{split}
    &\chi_L = \frac{\mathcal{L}_L\left(1+g_{LA}k_0^2V_A\mathcal{L}_A\right)}{1-g_{LA}g_{AL}k_0^4V_LV_A\mathcal{L}_L\mathcal{L}_A} \\
    &\chi_A = \frac{\mathcal{L}_A\left(1+g_{AL}k_0^2V_L\mathcal{L}_L\right)}{1-g_{LA}g_{AL}k_0^4V_LV_A\mathcal{L}_L\mathcal{L}_A}
\end{split}
\label{eq7}
\end{align}
where $k_0=\omega/c$. The complete solution to the effective susceptibility $\chi_{\text{eff}}$ is given by plugging Eq.~\ref{eq7} in Eq.~\ref{eq5}. 

Further simplifications are possible if we neglect the terms containing products of Green's functions. Then, $\chi_{\text{eff}}$ can be approximated by Eq.~\ref{eq6}.

\begin{align}
\begin{split}
    &\chi_{\text{eff}} =f_{L}\mathcal{L}_L+f_A\mathcal{L}_A+\frac{V_LV_A}{V_{\text{unit}}}k_0^2(h_{LA}+h_{AL})\mathcal{L}_L\mathcal{L}_A\times\\
    &\Bigg(1+\frac{2}{3}\text{cos}\left(\sqrt{3}k_0a\text{sin}\theta\text{cos}\phi\right)+ 2\text{cos}\left(k_0a\text{sin}\theta\text{sin}\phi\right)+\\
    &\frac{4}{3}\text{cos}\left(\frac{\sqrt{3}k_0a}{2}\text{sin}\theta\text{cos}\phi\right)\text{cos}\left(\frac{3k_0a}{2}\text{sin}\theta\text{sin}\phi\right) + \\
    &4\text{cos}\left(\frac{\sqrt{3}k_0a}{2}\text{sin}\theta\text{cos}\phi\right)\text{cos}\left(\frac{k_0a}{2}\text{sin}\theta\text{sin}\phi\right)\Bigg)\label{eq6}
\end{split}
\end{align}
where $f$ denotes the volume filling factor, which can be deduced from summing up the volume of individual domains and dividing over the unit cell volume, and $a$ is the smallest distance between the centers of A-stacking and L-stacking domains. The direction of the incident light is assumed to be $(\theta,\phi)$ represented in spherical coordinates (see Fig.~\ref{fig:panel2}a). The summation in Eq.~\ref{eq6} is taken over all the first-order neighbors. We use $h_{\text{AL(LA)}}$ to denote the Dyadic Green's function components isolated from complex exponential terms in Eq.~4-7 in $g_{\text{AL(LA)}}$ and cosine terms in Eq.~\ref{eq5}. 

The expression in Eq.~\ref{eq6} indicates the susceptibility is a function of incident light direction. In other words, the optical response of 1T-TaS$_2$ is non-local. When $k_0\cdot a<1$, the expression in Eq.~\ref{eq6} may be simplified further by approximating cosines with the first two terms of its Taylor series. This leads to the equation for non-local susceptibility as: 
\begin{align}
    \chi_{\text{eff}} \approx f_{L}\mathcal{L}_L+f_A\mathcal{L}_A+B\mathcal{L}_L\mathcal{L}_A\left(1-(\beta a)^2k_0^2\text{sin}\theta^2\right)
    \label{eq9}
\end{align}
\noindent where $(\beta a)$ is called the non-local parameter. $\beta$ is a constant that depends on the lattice geometry. $B$ is a dimensionless parameter proportional to $V_LV_Ak_0^2(h_{LA}+h_{AL})/V_{\text{unit}}$.
%For 1T structure of 1T-TaS$_2$, $\beta=\sqrt{3/8}$. 
Note that Eq.~\ref{eq9} is independent of $\phi$ because we assume that the domains in any layer of 1T-TaS$_2$ are randomly aligned and hence smooth out any anisotropy. 

%Nonlocality is characterized by $\beta$ from Taylor expansion coefficients. It is theorized to happen to oblique incidence with the incident angle $\theta$. Optical nonlocality is important only if $a/\lambda$ is smaller than one and cannot be negligibly small. 
The expression for non-local susceptibility in Eq.~\ref{eq9} is simple enough to employ it well in experimental studies. Plus, it still retains the key parameters that govern the physics of multi-phased quantum materials. For example, extracting the non-local parameter, and hence, $a$ can provide information on the competition between the relevant quantum phases. We demonstrate such utility of Eq.~\ref{eq9} by carrying out optical measurements on 1T-TaS$_2$ films at room temperature. The next section describes the non-local optical response of 1T-TaS$_2$ and the corresponding evolution of CDW stacking order with illumination intensity.

\section{Experimental results}
\subsection{Non-locality from angle-resolved reflectance measurements}
\begin{figure}
\centering
\includegraphics{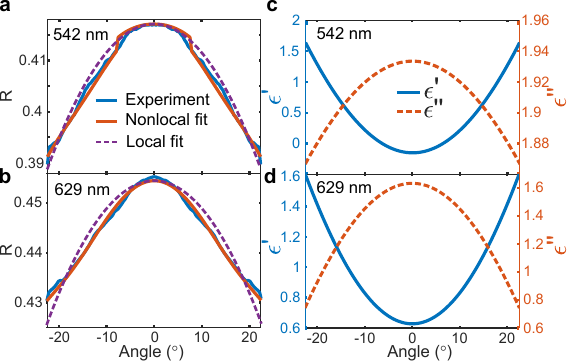}
\caption{\label{fig:panel3} Angle-resolved reflectance of 1T-TaS$_2$ film at an illumination wavelength of (a) 542 nm and (b) 629 nm. The experimental data (blue solid) are plotted against non-local model fit (orange solid) and local model fit (purple dashed). The real (blue solid) and imaginary (red dashed) parts of the extracted non-local dielectric function for an illumination wavelength of (c) 542 nm and (d) 629 nm. }
\end{figure} 

\begin{figure*}
\centering
\includegraphics{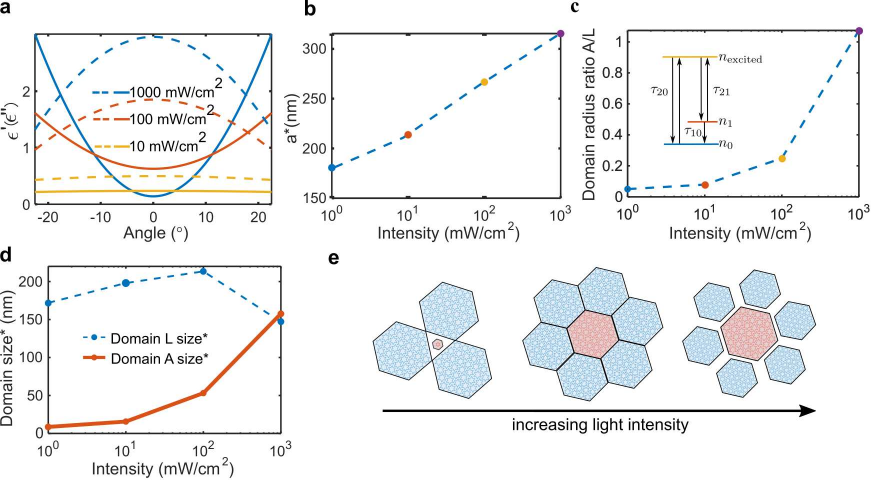}
\caption{\label{fig:panel4} Evolution of CDW stacking order with incoherent white light illumination intensity: (a) Non-local permittivity (real part in solid lines and imaginary part in dashed lines) extracted at three intensities of 628 nm wavelength light. (b) The inter-domain distance extracted from the non-local parameter at four distinct intensity values. (c) The ratio of the mean size of domains with A-stacking to that of L-stacking, called Domain radius ratio A/L, plotted against illumination intensity. (d) Average domain size for L-stacking (blue dashed) and A-stacking (red solid) versus illumination intensity. (d) An illustration of the inferred evolution of stacking domains in 1T-Ta$S_2$ with increasing illumination intensity. Blue and red hexagons represent the domains with stacking type L and A, respectively.} 
\end{figure*}

We used a Fourier-imaging spectroscopic microscope to measure the reflectance spectra on small area samples of 1T-Ta$S_2$. The small area samples of 1T-TaS$_2$ were prepared by mechanical exfoliation of bulk crystal purchased from commercial vendors that grow 1T-TaS$_2$ via the chemical vapor transport (CVT) method. The exfoliated flakes with a thickness of about 100 nm were chosen for this study. The flakes of 1T-TaS$_2$ used in this work were chemically stable in ambient air throughout this study. A white light LED with a maximum intensity of 1000 $\rm{mW/cm}^2$ and a central frequency of 650 nm was used for this experiment. TM polarization was used in our measurements. Angle-resolved reflectance spectra were acquired by inserting a Bertrand lens in the path of light exiting the microscope. More details on the setup and its calibration are described in our previous papers \cite{li2019plane,yang2022non}.

The measured angle-resolved reflectance was then fit with calculations employing a uniaxial permittivity model for 1T-TaS$_2$. The thickness of the 1T-TaS$_2$ sample was measured using an atomic force microscope (AFM). The dielectric function in the ab-plane was assumed to follow a Drude-Lorentz model. The c-axis permittivity was modeled by the non-local equation of Eq.~\ref{eq9}. For comparison, a separate fitting procedure was carried out on c-axis permittivity using a local Lorentz model. The fitting procedure was carried out first on the in-plane or ab-plane permittivity using the normal incidence reflectance. Then, the oblique incidence reflectance spectra were used to fit the c-axis permittivity. The reflectance data were fit simultaneously at all wavelengths and angular points. 

Fig.~\ref{fig:panel3}a and \ref{fig:panel3}b show a comparison between the local-model fit (Lorentz) and the non-local-model fit at two wavelengths in the visible. The errors in terms of standard deviation are 0.7 and 0.2, respectively. For our calibrated system, ordinary materials like silicon are benchmarked with $<0.01$ error. 
% The source of extra error includes the flatness of the sample, which is roughly $10$nm, and also the air-stability of 1T-TaS$_2$. Since the samples are far from the monolayer limit (~0.6nm), their exposure to air for one hour or should not cause any significant changes. 
Notably, the local model is a poor description of the c-axis optical properties of 1T-TaS$_2$. The fitting error sharply increases at higher angles of incidence for the local model. On the other hand, the non-local model for the c-axis dielectric constant of 1T-TaS$_2$ fits far better than the local model and serves as a good description of the room temperature optical properties of 1T-TaS$_2$ along the c-axis.
% Based on this trend of the error, we hypothesize that reflectance carries an extra dependence on the incident angle not captured by the Eigenmodes of the uniaxial light equation, and our nonlocal model provides a possible explanation that the finite sizes of the non-homogeneous domains cause this extra angular dependence.   
It is also interesting that our fitting results do not show any evidence of nonlocality along the in-plane direction. Only the c-axis permittivity $\epsilon_{zz}$ depends on the angle of incidence. We plot the non-local permittivity ($\epsilon'+i\epsilon''=1+\chi_{eff}$) of 1T-TaS$_2$ at two representative wavelengths in Fig.~\ref{fig:panel3}c and \ref{fig:panel3}d. The non-local dispersion is significant as can be seen from the plots. This is a consequence of large domain size or the value of $k_0a$ greater than unity.

% This result suggests in-plane homogeneity between $\alpha_{1xx}$ and $\alpha_{2xx}$. 

\subsection{Domain size dependence on light intensity}
The non-local model of Eq.~\ref{eq9} contains the non-local parameter $\beta a$. $\beta$ is a constant close to unity. Thus, extracting the non-local parameter from the fits provides information on the domain size. We use the information on domain size extracted via non-invasive single-shot optical measurement to elicit the underlying CDW physics in 1T-TaS$_2$. We performed angle-resolved reflectance measurements on the same flake of 1T-TaS$_2$ at different illumination intensities. Our previous work has shown that varying illumination intensity varies the CDW domain stacking preference. ~\cite{li2020large,li2021reorganization}. There is ample evidence of reflectance tunability on the order of $1-2$ percent when intensity is changed from 1000 mW/cm$^2$ to 10 mW/cm$^2$. While light-induced stacking reconfiguration is hypothesized as the mechanism of this tunability, there is no developed theory for this phenomenon. With our non-local optical characterization, we could probe the evolution of stacking preference together with the domain sizes as we sweep the illumination intensity. This information is key in developing a theory for the low-intensity light-matter interaction in 1T-TaS$_2$.

From angle-resolved reflectance measurements at three different wavelengths, the non-local permittivity is extracted at those same wavelengths. The non-local permittivity at a wavelength of 628 nm is shown in Fig.~\ref{fig:panel4}(a). The extracted permittivity value agrees well with previous reports~\cite{li2021reorganization,li2020large}. As the illumination gets more intense, 1T-TaS$_2$ becomes more absorptive. This is consistent with the previous observations that attributed the phenomenon to the increasing preference for A-stacking with intensity~ \cite{li2020large,li2021reorganization}. Since A-stacking domains are more metallic, the absorption increases~\cite{ritschel2018stacking}.

Next, we extracted the inter-domain distance $a*$ as a function of intensity, as shown in Fig.~\ref{fig:panel4}(b). Note that $a*$ is equal to $\beta a$ of Eq.~\ref{eq9}. Since $\beta$ is close to unity, we assume that $a*$ is close to the true domain size. Contrasting previous speculations, the domains get bigger with illumination intensity. Both types of stacking domains may grow in size with a higher illumination intensity. Alternatively, only one of them may grow with illumination intensity while the other shrinks. To find out further, we determined the volume fractions of A and L stacking domains from a three-level model as described in our previous work \cite{li2020large}. The three-level model is represented in the inset of Fig.~\ref{fig:panel4}c. The two stacking orders, L and A, are represented by states 0 and 1, respectively.  Since the energy difference between the two stacking orders is about 0.2 eV, much smaller than the energy of the illuminating photon, we hypothesize that the light-induced CDW stacking change in 1T-TaS$_2$ occurs via a high energy-intermediate state, called state 2 in the three-level model. The time constants $\tau$ of various transitions between the levels are shown in the inset of Fig.~\ref{fig:panel3}c. The steady-state population fraction of A-stacking $n_1$ is given by Eq.~\ref{eq10}.
\begin{equation}
n_1 = \frac{R\tau_{10}}{1+R\tau_{10}}\frac{R\tau_{20}\tau_{21}}{R\tau_{20}\tau_{21}+\tau_{20}+\tau_{21}}
\label{eq10}
\end{equation}

\noindent where $R$ is the photon absorption rate determined by the incident optical intensity and the absorption coefficient of 1T-TaS$_2$ film. Using the values of time constants from our previous work on time-resolved measurements on 1T-TaS$_2$ \cite{li2020large}, $\tau_{21}=\tau_{20}=3$ ns and $\tau_{10}=380$ ns. With these numbers, the L-stacking population fractions are 99.5, 98.8, 89.0, and 63.4 for illumination intensities of 1 mW/cm$^2$, 10 mW/cm$^2$,100 mW/cm$^2$,100 mW/cm$^2$, respectively. Note that the population fraction considers all CDW domains, not the uncondensed regions in the nearly commensurate phase. Using the population fraction information, we computed the ratio of the average domain size for stacking-A to that of stacking-L as shown in Fig.~\ref{fig:panel3}c. This computation was based on the geometry of a super-cell formed by close-packing A and L domains along with a domain boundary of an uncondensed metallic phase. The ratio of domain radius for A-stacking to L-stacking, called domain radius ratio A/L, increases with increasing illumination intensity.

Synthesizing the intensity dependence of domain size and domain radius ratio A/L together, the evolution of the sizes of A and L stacking domains with intensity may be computed. Fig.~\ref{fig:panel4}d shows that the size of A stacking domain increases over 500$\%$ as the illumination intensity increases from 1 to 1000 mW/cm$^2$. With light intensity increasing, A stacking domains grow at first to fill the uncondensed metallic spaces between the domains. Thus, the L-domain size remains almost constant in this regime. When the A-domains and L-domains are of the same size, the metallic regions in between domains occupy the smallest area. Further, increasing intensity causes the A-domains to grow bigger than L-domains by allowing larger uncondensed regions to fill the space in between (see  Fig.~\ref{fig:panel4}e). Here, the L-domains shrink in size to allow bigger A-domains and metallic inter-domain regions. This phenomenon might happen at intensities greater than 100 mW/cm$^2$. More experimental data points are necessary to verify our hypothesis in this intensity range.

\section{Conclusions}
We developed a non-local model to describe the low-intensity light-matter interaction in 1T-TaS$_2$. Our model was based on self-consistently solving for local fields in the two domains with different CDW stacking orders. We verified the non-local model via experimental measurements of angle-resolved reflectance on a mechanically exfoliated flake of 1T-TaS$_2$. The non-local model was not only a more precise description of the optical properties of 1T-TaS$_2$ but also a non-invasive probe of the CDW domain size. Using the information on domain sizes at various illumination intensities extracted from the non-local model, a mechanism for the light-induced reorganization of CDW stacking order in 1T-TaS$_2$ was possible. The non-local model described here and its application on 1T-TaS$_2$ shown here may be extended to any other quantum material with inhomogeneity. Non-locality would non-invasively map the domain size and hence the competition between quantum phases under various conditions. Thus, the non-local model developed here, together with the non-linear model previously developed by our group, could describe the low-intensity optical properties of all multi-phased quantum materials and serve as a powerful non-invasive tool to probe the energy landscape of strong correlations in these materials.

\end{document}